# New high-pressure phase of HfTiO$_4$ and ZrTiO$_4$ ceramics

Daniel Errandonea[1,†,*], David Santamaria-Perez[2,*], Tatiana Bondarenko[3], and Oleg Khyzhun[3]

[1] Departamento de Física Aplicada-ICMUV, Universitat de València, Edificio de Investigación, c/Dr. Moliner 50, 46100 Burjassot (Valencia), Spain.

[2] Departamento de Química Física I, Universidad Complutense de Madrid, Avda. Complutense s/n, 28040 Madrid, Spain

[3] Frantsevych Institute for Problems of Materials Science, National Academy of Sciences of Ukraine, 3 Krzhyzhanivsky Street, UA-03142 Kyiv, Ukraine

**Abstract:** We studied the high-pressure effects on the crystalline structure of monoclinic HfTiO$_4$ and ZrTiO$_4$. We found that the compressibility of these ceramics is highly non-isotropic, being the *b*-axis the most compressible one. In addition, the *a*-axis is found to have a small and negative compressibility. At 2.7 GPa (10.7 GPa) we discovered the onset of a structural phase transition in HfTiO$_4$ (ZrTiO$_4$), coexisting the low- and high-pressure phases in a broad pressure range. The new high-pressure phase has a monoclinic structure which involves an increase in the Ti-O coordination and a collapse of the cell volume. The equation of state for the low-pressure phase is also determined.

---

[†] electronic mail: daniel.errandonea@uv.es, Tel.: (34) 96 354 4475, FAX: (34) 96 3543146
[*] Member of MALTA Consolider Team

# I. Introduction

Microwave technology has made remarkable progress, along with development of advanced communications systems that include global positional systems. This technology needs the use of stable high-dielectric ceramics. $HfTiO_4$ and $ZrTiO_4$ are used for these applications and therefore their structural and electrical properties at ambient pressure have been widely studied [1, 2]. It is known that $HfTiO_4$ and $ZrTiO_4$ have a columbite-type ($\alpha$-$PbO_2$) structure. It has orthorhombic symmetry, space group *Pbcn*, with the Hf (Zr) and Ti atoms randomly distributed on the octahedral site of the $\alpha$-$PbO_2$ structure [3]. The temperature effects on the crystalline structure of $ZrTiO_4$ have also been studied [4]. Upon cooling it undergoes a lattice contraction and a cation-positional ordering. It is also known that the application of an electric field affects these structural changes [5]. On the contrary nothing is known yet about how the application of pressure affects the structural properties of $HfTiO_4$ and $ZrTiO_4$. This information is relevant in several areas of materials research and it necessary to better understand their mechanical properties. The importance of high-pressure structural studies has been shown in the case of related $ABO_4$ oxides; e.g. arsenates [6], germanates [7], molybdates [8], phosphates [9], tungstates [10], and vanadates [11]. In order to shed some light on the understanding of the mechanical properties of titanate ceramics, we report high-pressure x-ray diffraction studies of $HfTiO_4$ and $ZrTiO_4$ up to a pressure of 17 GPa. The bulk and axial compressibilities of their structures are reported. In addition, we report by the first time the finding in both titanates of a pressure-induced phase transition to a monoclinic structure. The present work contributes to achieve a deeper understanding of pressure effects on the crystal structure of $ABO_4$ oxides of technological importance [12].

**II. Experimental details**

Ceramics with the ZrTiO$_4$ and HfTiO$_4$ compositions were prepared by the conventional mixed metal-oxide method. Briefly, stoichiometric molar mixtures of TiO$_2$ and ZrO$_2$ (HfO$_2$) powders were pressed at 180 bar to produce pellets with diameter ~8 mm and height ~4 mm. The pellets were heated to 1300 °C and calcinated in air at the mentioned temperature for 6 h. After regrinding and repressing, in order to attain a better homogeneity of the products obtained, a second heating treatment at 1700 °C was applied for 24 h. The samples used in our experiments were pre-pressed pellets prepared using a finely ground powder obtained from the synthesized titanates. Before loading the samples into the diamond-anvil-cell (DAC), their phase purity was tested with x-ray diffraction. The diffraction lines for HfTiO$_4$ (ZrTiO$_4$) collected at ambient pressure (0.0001 GPa = 1 bar) showed only a single columbite-type phase with unit-cell parameters $a$ = 4.746 Å, $b$ = 5.558 Å, and $c$ = 5.038 Å ($a$ = 4.795 Å, $b$ = 5.444 Å, and $c$ = 5.022 Å), in agreement with reported literature values [13, 14].

High-pressure angle-dispersive x-ray diffraction (ADXRD) measurements on HfTiO$_4$ and ZrTiO$_4$ have been carried out with an Xcalibur diffractometer (Oxford Diffraction Limited). X-ray diffraction patterns were obtained on a 135-mm Atlas CCD detector placed at 110 mm from the sample using $K_{\alpha 1}$: $K_{\alpha 2}$ molybdenum radiation. The x-ray beam was collimated to a diameter of 300 μm. High-pressure measurements on HfTiO$_4$ and ZrTiO$_4$ were performed at room temperature (RT) in a modified Merrill-Bassett diamond-anvil cell (DAC) up to 9 and 17 GPa, respectively. The diamond anvils used have 500-μm-size culet. The titanates powders were placed in the 150-μm-diameter hole of the stain-steel gasket preindented to a thickness of 60 μm. A 4:1 methanol:ethanol mixture was used as quasi-hydrostatic pressure-transmitting medium [15, 16] and ruby chips evenly distributed in the pressure chamber were used to measure the pressure by the ruby fluorescence method [17]. Exposure times were typically of 80 minutes. The DAC used for the experiments allows access to an angular range $2\theta = 25º$. An exposure on the starting material, at room conditions, in a 0.3-mm glass capillary was obtained using the same installation with the sample to CCD distance of 110 mm. The observed intensities were integrated as a function of $2\theta$ in order to give conventional one-dimensional diffraction profiles. The CrysAlis software, version 171.33.55 (Oxford Diffraction Limited), was used for the data collections and the preliminary reduction of the data. The indexing and refinement of the powder patterns were performed using the POWDERCELL [18] and FULLPROF [19] program packages.

**III. Results and Discussion**

The in situ ADXRD data of $HfTiO_4$ measured at different pressures are shown in Fig. 1. The x-ray patterns could be indexed with the orthorhombic structure (phase I), stable at normal conditions, up to 2.7 GPa. At this pressure the appearance of an additional peak (depicted by an arrow) is seen near $2\theta = 13º$. Upon further compression additional peaks appears and the diffraction peaks of the low-pressure phase gradually lost intensity. These facts indicate the onset of a phase transition to a phase we will denote as phase II. The transition is completed at 8.4 GPa and is not fully reversible upon decompression. The diffraction patterns of the high-pressure phase suggest a decrease of the crystal symmetry. The most distinctive reflections of phase II are those depicted by arrows in the figure. Another distinctive feature of the high-pressure phase is the broadening and splitting of the peak located around $2\theta = 11º$. From the diffraction patterns collected at different pressures we extracted the pressure evolution of the lattice parameters of orthorhombic $HfTiO_4$. We also propose a crystalline structure for the high-pressure phase. The x-ray diffraction patterns are plotted together with the refined structure model to illustrate the quality of the structural refinements. The obtained evolution for the unit-cell parameters and the cell volume are shown in Fig. 2. There it can be seen that the most compressible axis is the *b*-axis. In addition, the *a*-axis is nearly uncompressible, showing a tendency to slightly increase upon compression. As a consequence of these facts the anisotropy of the crystal is reduced by pressure. This is basically visualized by the merging under compression of the (021) and (200) Bragg peaks. The pressure–volume curve shown in Fig. 2 was analyzed in the standard way using a third-order Birch–Murnaghan equation of states (EOS) [20]. The bulk modulus ($B_0$), its pressure derivative ($B_0$'), and the atomic volume ($V_0$) at zero pressure obtained for the low-pressure phase are $B_0$ = 147(5) GPa, $B_0$' = 4.5(9), and $V_0$ = 133.1(5) Å$^3$. The fitted EOS is shown as a solid line in Fig. 2.

Since the low-pressure phase of $ZrTiO_4$ and $HfTiO_4$ can be viewed as a superstructure of $ZrO_2$ [19], in order to propose a crystalline structure for the high-pressure of $HfTiO_4$ we have considered all the known phases of $ZrO_2$ and $TiO_2$ [20, 21]. Among these structures, the cubic pyrite and fluorite, tetragonal anatase and rutile, and orthorhombic cotunnite cannot explain all the Bragg peaks of the high-pressure phase. In contrast we can explain them considering a monoclinic baddeleyite-type structure (space group *P2$_1$/c*); see Fig. 1. For this structure at 8.65 GPa we obtain the following unit-cell parameters: *a* = 4.838(8) Å, *b* = 4.945(8) Å, *c* = 5.060(9) Å, and β = 98.2(2)º. On it the Hf and Ti atoms are randomly distributed at Wyckoff position 4e (0.2759, 0.0395, 0.2086). Both cations are seven coordinated by oxygen atoms. This structure implies a volume collapse of 5% at the transition and an increase of the Ti-O coordination from 6 to 7. The orthorhombic-monoclinic transition is of first-order type. Both facts are consistent with the high-pressure behaviour of related oxides [16, 22].

The proposed transition is also consistent with the rutile-columbite-baddeleyite high-pressure structural sequence of $TiO_2$ and isomorphic oxides [24].

As we mentioned above there is a 5 GPa coexistence region between phases I and II. The percentage of each of the two phases present in the sample can be estimated from the relative intensity of the main peaks of phases I and II that do not overlap [25]. Fig. 3 shows the relative fractions of monoclinic $HfTiO_4$ as estimated from $I_{II} / (I_I + I_{II})$, where $I_I$ and $I_{II}$ are the integrated area of the (111) reflection of phase I and the ($\bar{1}11$) reflection of phase II, respectively. These data show that the monoclinic/orthorhombic ratio gradually increases with pressure, reaching 1 at 8.4 GPa when the transition is completed.

Fig. 4 shows a selection of the diffraction patterns we collected for $ZrTiO_4$. Qualitatively, this ceramic material follows the same trend under compression that $HfTiO_4$. In this case new Bragg peaks appear at 10.7 GPa. Upon further compression additional peaks appear (depicted by arrows in the figure), other peaks broaden, and the peaks of the low-pressure phase gradually lost intensity. Clearly, the onset of a phase transition takes place at 10.7 GPa. The high-pressure phase can be explained by the same baddeleyite-type structure that describes the high-pressure phase of $HfTiO_4$. For this structure at 14.2 GPa we obtain the following unit-cell parameters: $a$ = 4.836(8) Å, $b$ = 4.995(8) Å, $c$ = 5.075(9) Å, and β = 99.0(2)°. This structure implies a volume collapse of 2% at the transition. It explains all the new diffraction peaks present in the diffraction pattern and also the broadening of the (110) peak of phase I, which occurs because of the splitting of the (011) and (110) peaks of phase II. On decompression we have found that the phase transition is reversible, but the low-pressure phase is fully recovered only below 6.9 GPa. This large hysteresis and the volume collapse observed at the transition indicate that this has a first-order character. The hysteresis observed in ZrTiO$_4$ also explain why the transition in not fully reversible in HfTiO$_4$, in which the transition onset is at a lower pressure (2.7 GPa).

The obtained evolution for the unit-cell parameters and the cell volume for ZrTiO$_4$ is shown in Fig. 5. There it can be seen that, as happens in HfTiO$_4$, the most compressible axis is the *b*-axis. Again, the *a*-axis is nearly uncompressible, showing a tendency to slightly increase upon compression. As a consequence of these facts at 16 GPa the difference between the *a*- and *c*-axes is smaller than 2%, i.e. the crystal under compression becomes pseudo-tetragonal. The pressure–volume curve shown in Fig. 5 was analyzed in the standard way using a third-order Birch–Murnaghan equation of states (EOS) [18]. The bulk modulus ($B_0$), its pressure derivative ($B_0'$), and the atomic volume ($V_0$) at zero pressure obtained for the low-pressure phase are $B_0$ = 144(5) GPa, $B_0'$ = 3.6(9), and $V_0$ = 131.1(5) Å$^3$. The fitted EOS is shown as a solid line in Fig. 5. It is important to note that both titanates have a quite similar bulk modulus. This parameter is 40% smaller than that of columbite-type HfO$_2$, TiO$_2$, and ZrO$_2$ [22, 23, 26], suggesting that the disorder of cations in HfTiO$_4$ and ZrTiO$_4$ could favour the volume reduction of the crystal under compression.

Regarding the coexistence of phases I and II in ZrTiO$_4$, we have applied to this ceramic the same criterion that in HfTiO$_4$ to determine the proportion of the high-pressure phase present at different pressures. In Fig. 3 it can be seen that both materials follows the same trend with the only difference that the transition onset is at lower pressure in HfTiO$_4$. The extrapolation of the data we have for ZrTiO$_4$ suggests that for this material the transition will be completed at 20 GPa.

## IV. Summary

We have studied the structural properties of $HfTiO_4$ and $ZrTiO_4$ under high-pressure. We found that both ceramics follow a qualitatively similar behaviour. The compression of the low-pressure phase is highly anisotropic being the *b*-axis the most compressible one. In addition, compression induces a first-order transition from the low-pressure columbite-type structure to a high-pressure baddeleyite-type structure. The transition involves a volume collapse, a Ti-O coordination increase, and a decrease of the crystal symmetry from orthorhombic to monoclinic. Finally, the EOS for the low-pressure phase is determined being the bulk modulus of $HfTiO_4$ ($ZrTiO_4$) 147 (144) GPa.

**Acknowledgements:** Financial support from Spanish Consolider Ingenio 2010 Program (Project No. CSD2007-00045) is acknowledged. This work was also supported by Spanish MICCIN (Grant No. MAT2007-65990-C03-01).

**Figure Captions**

**Figure 1:** Selection of RT ADXRD data of HfTiO$_4$ at different pressures up to 8.65 GPa. In all diagrams, the background was subtracted. Pressures are indicated in the plot. In the ADXRD patterns at 1 bar and 8.65 GPa we show the refined profile (dotted lines). The bars and Miller indexes indicate the calculated positions of the Bragg reflections.

**Figure 2:** Pressure dependence of the unit-cell parameters and volume in orthorhombic HfTiO$_4$. In the upper plot the solid line represents the EOS fitted. In the lower plot the lines are a guide to the eye.

**Figure 3:** Relative amount of monoclinic phase I to orthorhombic phase II.

**Figure 4:** Selection of RT ADXRD data of ZrTiO$_4$ at different pressures up to 14.2 GPa. In all diagrams, the background was subtracted. Pressures are indicated in the plot. In the lowest trace the peaks of the orthorhombic phase are labeled. Most distinctive peaks of the monoclinic phase are indicated at 11.4 GPa.

**Figure 5:** Pressure dependence of the unit-cell parameters and volume in orthorhombic ZrTiO$_4$. In the upper plot the solid line represents the EOS fitted. In the lower plot the lines are a guide to the eye.

**Figure 1**

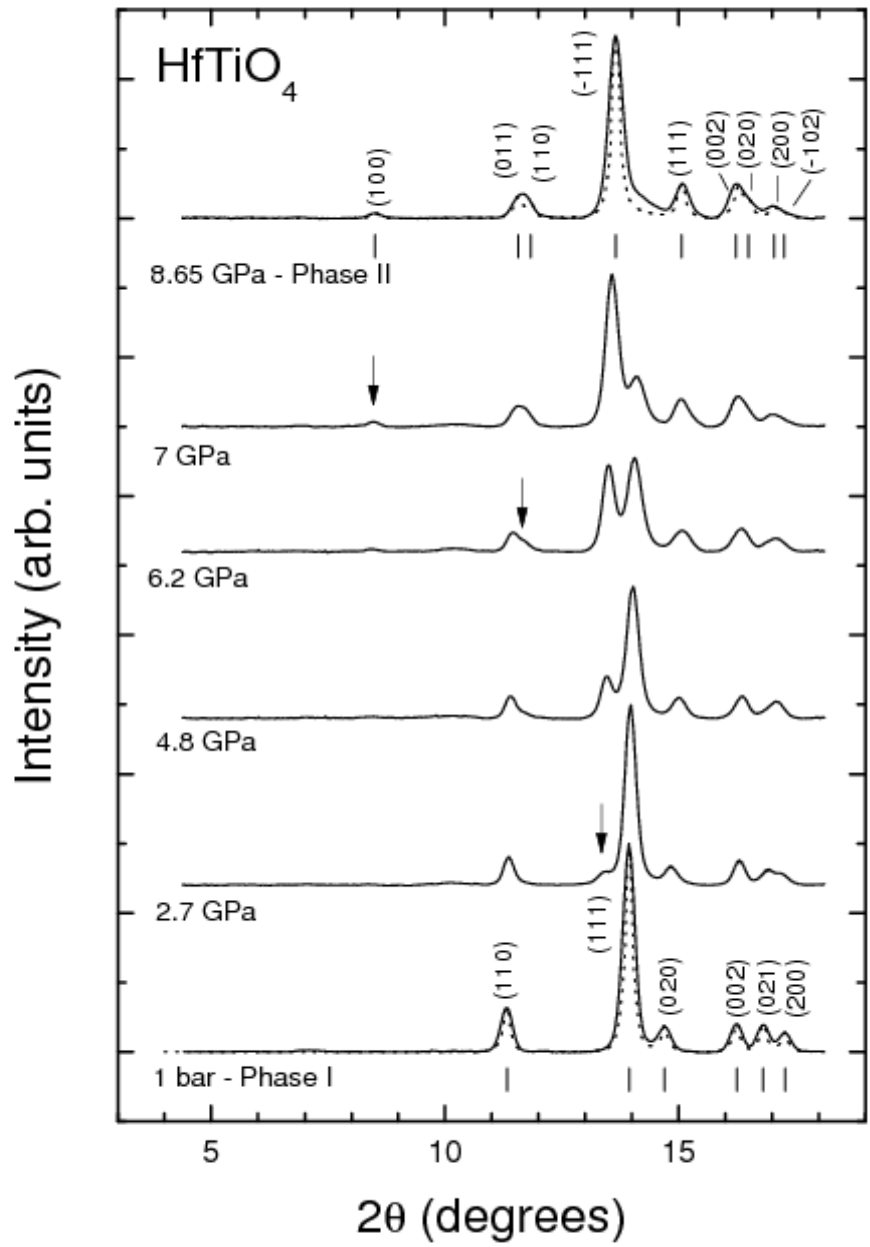

**Figure 2**

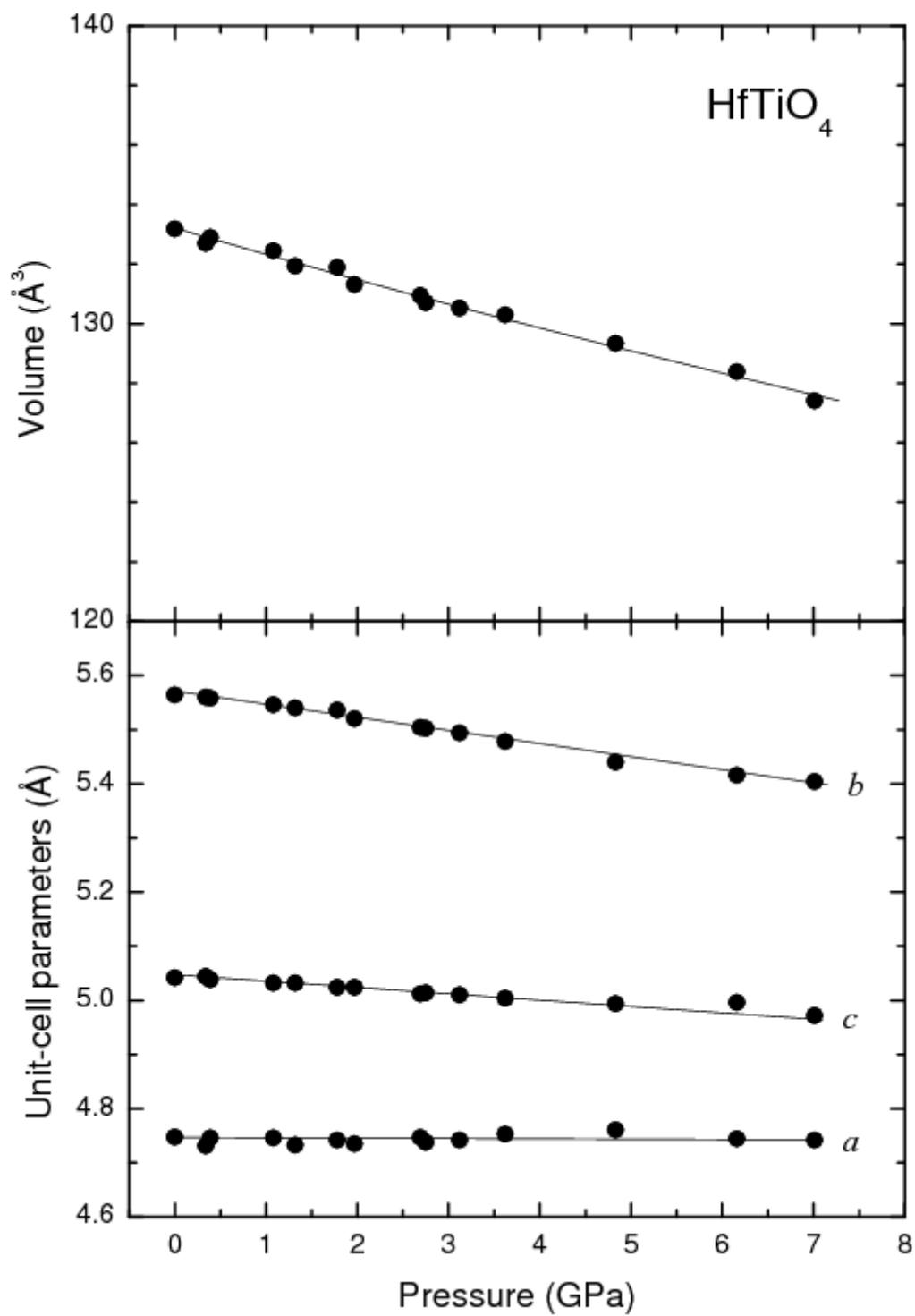

**Figure 3**

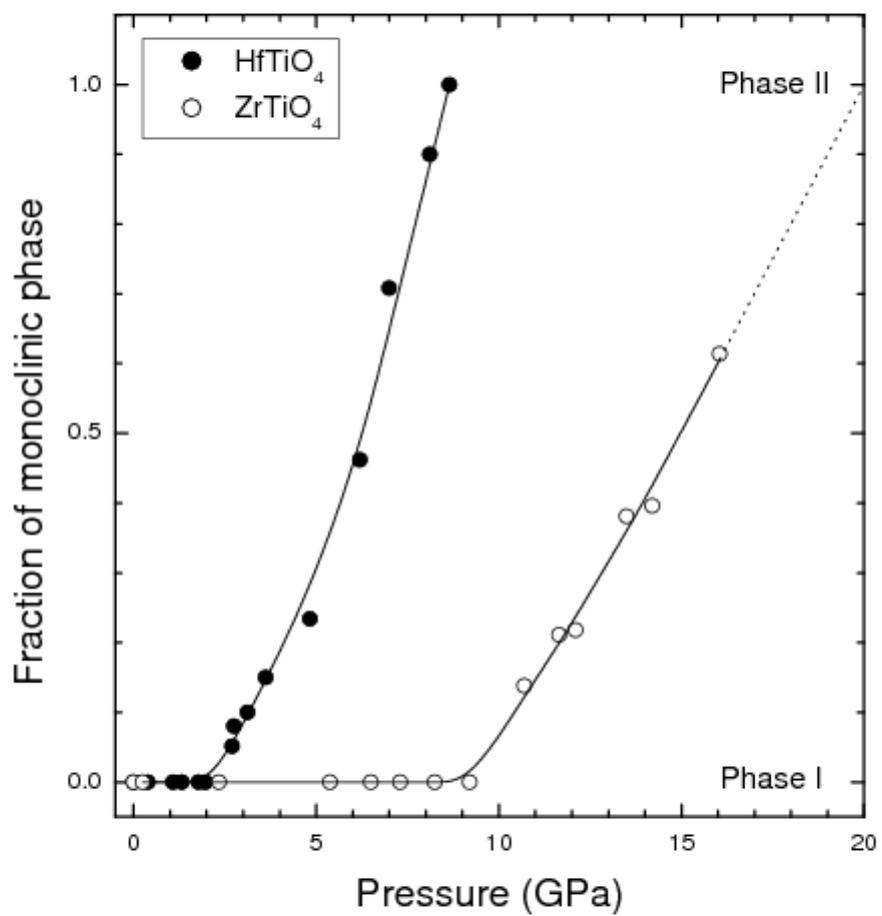

**Figure 4**

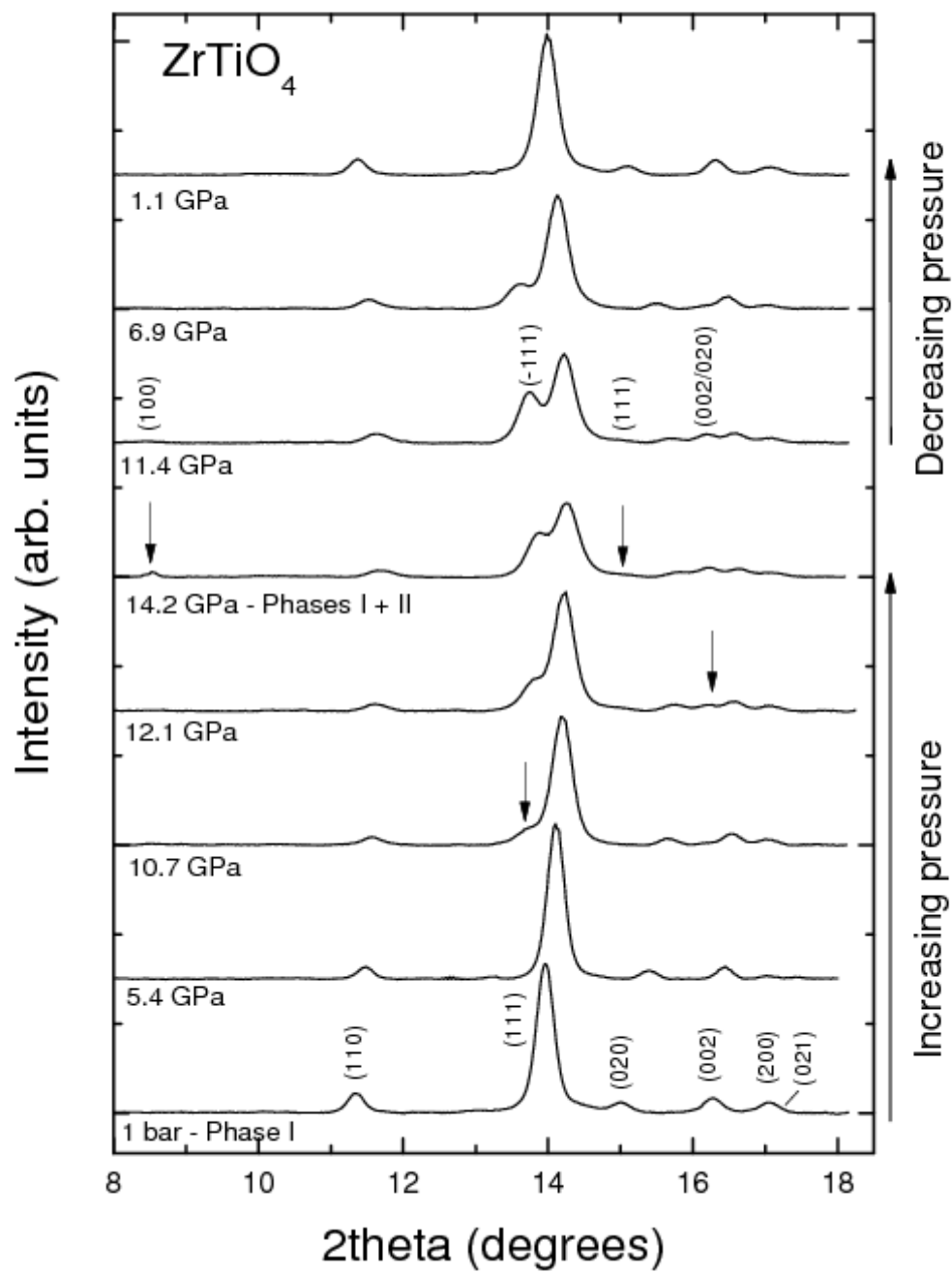

**Figure 5**

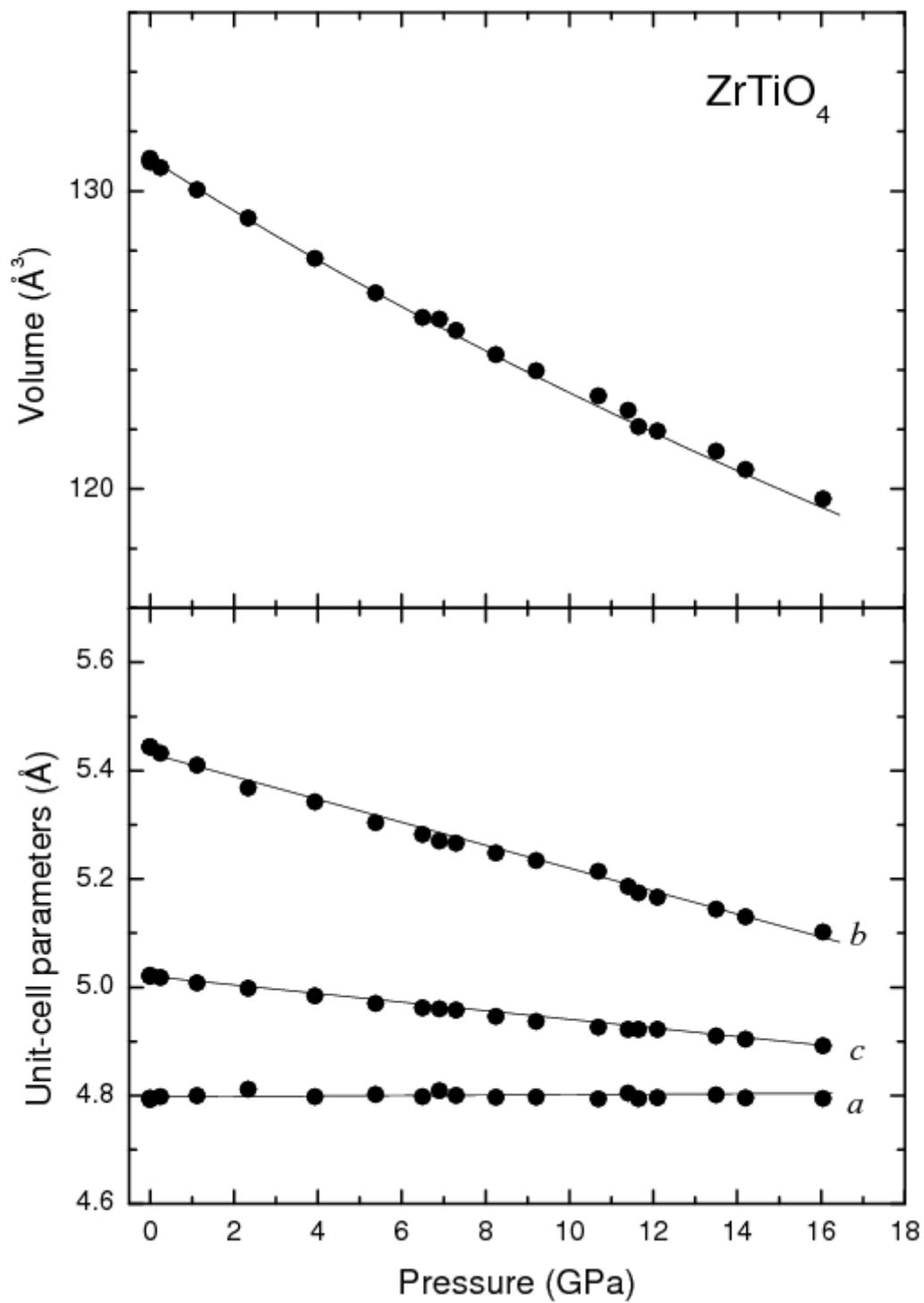